\documentclass[prl,showpacs,floatfix,aps]{revtex4}
\usepackage{hyperref}
\usepackage{amssymb}
\usepackage{amsmath}
\usepackage{graphicx,bm}
\usepackage[utf8]{inputenc}
\usepackage[T1]{fontenc}
\usepackage{textcomp}

\hypersetup{colorlinks = true, linkcolor = blue, citecolor = blue}
\begin{document}

\title{\textbf{Evolution of the pseudogap temperature dependence in YBa$_2$Cu$_3$O$_{7-\delta}$ films under the influence of a magnetic field}}

\author{E.\,V.\,Petrenko$^{1}$, A.\,V.\,Terekhov$^{1}$, L.\,V.\,Bludova$^{1}$, Yu.\,A.\,Kolesnichenko$^{1}$, N.\,V.\,Shytov$^{1}$, D.\,M.\,Sergeyev$^{2}$, E.\,L{\"a}hderanta$^{3}$, K.\,Rogacki$^{4}$, A.\,L.\,Solovjov$^{1,3,4}$}

\affiliation{$^1$B.\,Verkin Institute for Low Temperature Physics and Engineering of National Academy of Science of Ukraine, Kharkiv 61103, Ukraine\\
$^2$K.Zhubanov Aktobe Regional State University, 030000 Aktobe, Kazakhstan\\
$^3$Lappeenranta University of Technology, School of Engineering Science, 53850 Lappeenranta, Finland\\
$^4$Institute of Low Temperature and Structure Research of Polish Academy of Science, 50-422 Wroclaw, Poland\\
} E-mail: petrenko@ilt.kharkov.ua

\begin{abstract}
The evolution of the temperature dependence of pseudogap
$\Delta$*(T) in optimally doped (OD) YBa$_2$Cu$_3$O$_{7-\delta}$
(YBCO) films with $T_{c}$ = 88.7 K under the influence of a magnetic
field $B$ up to 8 T has been studied in detail. It has been
established that the shape of $\Delta$*(T) for various $B$ over the
entire range from the pseudogap opening temperature $T$* to
$T_{01}$, below which superconducting fluctuations occur, has a wide
maximum at the BEC-BCS crossover temperature $T_{pair}$, which is
typical for OD films and untwinned YBCO single crystals. $T$* was
shown to be independent on $B$, whereas $T_{pair}$ shifts to the low
temperature region along with increase of $B$, while the maximum
value of $\Delta$*($T_{pair}$) remains practically constant
regardless of $B$. It was revealed that as the field increases, the
low-temperature maximum near the 3D-2D transition temperature
$T_{0}$ is blurred and disappears at $B$ > 5 T. Moreover, above the
Ginzburg temperature $T_{G}$, which limits superconducting
fluctuations from below, at $B$ > 0.5 T, a minimum appears on
$\Delta$*(T) at $T_{min}$, which becomes very pronounced with a
further increase in the field. As a result, the overall value of
$\Delta$*(T) decreases noticeably most likely due to pair-breaking
affect of a magnetic field. A comparison of $\Delta$*(T) near
$T_{c}$ with the Peters-Bauer theory shows that the density of
fluctuating Cooper  pairs actually decreases from \textless
n$_{\uparrow}$n$_{\downarrow}$\textgreater\hspace{0pt} $\approx$
0.31 at $B$ = 0 to \textless
n$_{\uparrow}$n$_{\downarrow}$\textgreater\hspace{0pt} $\approx$
0.28 in the field 8 T. The observed behavior of $\Delta$*(T) around
$T_{min}$ is assumed to be due to the influence of a two-dimensional
vortex lattice created by the magnetic field, which prevents the
formation of fluctuating Cooper pairs near $T_{c}$.


\end{abstract}

\maketitle

$\textbf{1. INTRODUCTION}$\\

The number of works devoted to high-temperature superconductors
(HTSCs) keeps growing steadily, and this is not surprising: after
the resonant publication of Lee $et$ $al.$ \cite{LK99}, interest in
such materials has increased even more. Unfortunately, so far no
scientific group in the world can confirm the sought-after
superconductivity at normal pressure\cite{Garisto, Puphal} declared
by \cite{LK99} in modified lead apatite. The impossibility of
synthesizing compounds with zero resistance at room temperatures and
ambient pressure is primarily due to the lack of a complete
understanding of the mechanism of superconducting pairing at
superconducting transition temperatures above 100 K.

In addition, cuprates at some temperature range above $T_{c}$
exhibit an unusual feature known as a pseudogap (PG) \cite{Kord,
Gao, Tallon, Peng, Chakraborty, Esterlis, Yu}, which opens in
lightly doped YBa$_{2}$Cu$_{3}$O$_{7-\delta}$ (YBCO) at a
characteristic temperature $T$* >> $T$$_{c}$ \cite{Loktev, Sol_rij}.
It is believed that studying the PG phenomenon can definitely shed
light on the microscopic mechanism of high-temperature
superconductivity \cite{Sol_rij, Kord} (and references there in).
However, the physics behind PG is still uncertain. The
aforementioned YBCO cuprates, as is known, belongs to the class of
metal oxides with active CuO$_{2}$ planes and, in addition to high
$T$$_{c}$ and PG, have a low charge carrier density $n$$_{f}$,
strong electronic correlations, quasi-two-dimensionality and, as a
consequence, strong anisotropy of electronic properties
\cite{Haussmann, Loktev, Tchernyshyov, Engel}. In particular, the
coherence length along the $ab$ plane, which determines the size of
Cooper pairs in HTSCs, is $\xi$$_{ab}$(T) $\approx$
10$\xi$$_{c}$(T), where $\xi$$_{c}$(T) is the coherence length along
the $c$-axis \cite{Sol_rij}.

We support an idea that a low charge carrier density, $n$$_{f}$, is
a necessary condition for the formation of paired fermions in
cuprates below the characteristic temperature $T$* >> $T$$_{c}$, the
so-called local pairs (LPs) \cite{Tchernyshyov, Emery}, which, most
likely, responsible for the formation of the PG state
(\cite{Sol_rij, Emery, Randeria, Sol-Rog2023} and references
therein). At high temperatures $T$ $\leq$ $T$*, LPs appear in the
form of strongly bound bosons (SBBs), which obey the Bose-Einstein
condensation theory (BEC) \cite{Randeria}. Since $\xi$$_{ab}$(T*)
$\approx$ 10 \AA\hspace{0pt} in YBCO, SBBs are small, but very
tightly bound pairs, since, according to the theory \cite{Haussmann,
Engel}, the binding energy in a pair $\epsilon$$_{B}$ $\sim$
1/$\xi_{ab}^2$. As a result, SBBs are not destroyed by thermal
fluctuations and any other influences. However, with decreasing
temperature, $\xi$$(T)$ and, consequently, the pair size increase,
and the SBBs gradually transform into fluctuating Cooper pairs
(FCPs), obeying the BCS theory in the range of superconducting (SC)
fluctuations near $T$$_{c}$ \cite{Emery, Randeria, Sol-Rog2023}. We
would like to emphasize that condensation of the FCPs into the SC
state is possible only from the three-dimensional (3D) state
\cite{Haussmann, Engel}. Therefore, when $\xi$$_{c}$(T) exceeds the
size $d$ of the YBCO unit cell along the c-axis: $\xi$$_{c}$(T) >
$d$ near $T$$_{c}$, the quasi-two-dimensional (2D) state of the
HTSCs always changes to the 3D state \cite{Sol_rij, SolDmReview}.

Nevertheless, some other models have been proposed to explain the
physics of PG, such as spin fluctuations \cite{Stojkovic}, charge
(CDW) \cite{Badoux, Gabovich} and spin (SDW) \cite{Taillefer,
Badoux, Peng} density waves, charge ordering (CO) (\cite{Badoux,
Peng, Dzhumanov1, Dzhumanov2} and references therein) and even pair
density waves (PDW) \cite{Chakraborty, Wang}. However, in spite of
the fact that the interest in the PG study has noticeably increased
in recent years \cite{Chakraborty, Wang, Kivelson, Robinson,
Mishra}, the physics of the PG state is still not completely clear.
According to the latest concepts \cite{Peng,Sol-Rog2023, Taillefer},
below $T$* a rearrangement of the Fermi surface is quite possible,
which largely determines the unusual properties of cuprates in the
PG region.  At the same time, despite the fact that the number of
works devoted to the study of HTSCs and, in particular, the PG, is
extremely large, there is a lack of papers studying the influence of
a magnetic field on excess conductivity in cuprates. However, it is
precisely the study of the influence of the magnetic field on
fluctuation conductivity (FLC) and PG that can answer the question:
which of the physical mechanisms of PG considered above actually
takes place in cuprates?

To the best of our knowledge, the effect of magnetic field on FLC in
YBCO materials has been analyzed in a fairly small number of studies
so far \cite{VovkExConMagF, Malik, Rey, PetrExConMagF}. However, in
\cite{VovkExConMagF} the applied field does not exceed 1.27 T, which
is too small to draw any firm conclusions. In \cite{Malik} magnetic
field $B$ = 12 T and accordingly in \cite{Rey} magnetic field $B$ =
9 T were used. But in \cite{Malik}, it is reported a decrease in
$\xi$$_{c}$(T) along with a decrease in $T_{c}$ with increasing
magnetic field. It is amazing, since in the traditional theory of
superconductivity $\xi$ $\sim$ 1/$T_c$ \cite{DeGennes}. In addition,
in both papers the authors did not show the evolution of the FLC
with a magnetic field. Moreover, we are not aware of any work that
has studied the effect of the magnetic field on the PG in HTSC's.

In our previous work \cite{PetrExConMagF}, we tried to shed light on
the mechanism of the influence of the magnetic field on FLC in YBCO
by studying the influence of the magnetic field in the $ab$-plane on
the resistivity $\rho$(T) and fluctuation conductivity $\sigma'$(T)
of a thin YBCO film at increasing the magnetic field to 8 T. The
corresponding dependences $\sigma'$(T) for all applied fields were
carefully analyzed. As expected, at $B$ = 0, $\sigma'$(T) near
$T$$_{c}$ was described by the 3D Aslamasov-Larkin (AL) theory
\cite{ODYBCO1GPa,AL} and the 2D Maki-Thompson (MT) fluctuation
theory \cite{HL} above the crossover temperature $T$$_{0}$. However,
at $B$ = 3 T, the MT term is completely suppressed, and above
$T$$_{0}$, $\sigma'$(T) is unexpectedly described by the 2D-AL
fluctuation contribution. At the same time, $\xi$$_{c}$(0) abruptly
increases, that is, demonstrates a rather unusual $\xi$$_{c}$(0) vs
$T$$_{c}$ dependence in magnetic field (Fig. 7, curve 1 in
\cite{PetrExConMagF}). However, the evolution of the pseudogap under
the influence of a magnetic field has not been studied.

In this paper, for the first time, a detailed analysis of the
temperature and field dependences of the pseudogap $\Delta$*(T,B) of
a thin YBCO film in a magnetic field of up to 8 T was carried out,
also using all the necessary parameters obtained from the analysis
of fluctuation conductivity $\sigma'$(T,B) in the above-mentioned
work. Thus, in a sense, the article is a logical continuation of our
previous research \cite{PetrExConMagF}. It was found that at $B$ = 0
the shape of $\Delta$*(T) was expectedly typical for optimally doped
films \cite{SolDmReview} and non-twinned single crystals
\cite{Sci.Rep.9-ybco_an}, but unexpectedly changed at $B$ = 8T.  It
was found that in the range of SC fluctuations below $T_{01}$, the
magnetic field strongly affects both the shape and the value of
$\Delta$*(T,B), which can be seen by comparing the corresponding
temperature dependences of the PG at different B near $T_{c}$. To
gain a deeper understanding of the influence of the magnetic field
on superconducting fluctuations, a comparison was made with the
Peters-Bauer theory \cite{PB} and the local pair density near
$T_{c}$ in the film under study was estimated as a function of the
applied field. A detailed discussion of the results obtained is given below.\\

$\textbf{2. EXPERIMENT}$\\

Epitaxial YBCO films were deposited at $T$ = 770\textdegree C and an
oxygen pressure of 3 mbar at
(LaAlO$_{3}$)$_{0.3}$(Sr$_{2}$TaAlO$_{6}$)$_{0.7}$ substrates, as
described elsewhere \cite{Przyslupski}. The thickness of deposited
films, d $\approx$ 100 nm, was controlled by the deposition time of
respective targets. X-ray analyses have shown that all samples are
excellent films with the $c$ axis perfectly oriented perpendicular
to the CuO$_{2}$ planes. Next, the films were lithographically
patterned and chemically etched into well-defined 2.35 $\times$ 1.24
mm$^2$ Hall-bar structures. To perform contacts, golden wires were
glued to the structure pads by using silver epoxy. Contact
resistance below 1 $\Omega$ was obtained. The main measurements
included a fully computerized setup, the Quantum Design Physical
Property Measurement System (PPMS-9), by using the AC current of
$\sim$ 100 $\mu$A at 19 Hz. The four-point probe technique was used
to measure the in-plane resistivity $\rho$$_{ab}$(T) = $\rho$(T). A
static magnetic field was created using a superconducting magnet.
Measurements were performed up to 9 T for the field orientations
parallel
to the $B$ || $ab$ . \\

$\textbf{3. RESULTS AND DISCUSSION}$\\

$\textbf{3.1. Resistivity}$\\

The temperature dependence of the resistivity $\rho$(T) of the
YBa$_2$Cu$_3$O$_{7-\delta}$ film in the absence of an external
magnetic field is shown in Fig. 1 and is typical for all underdoped
cuprates. For temperature above $T$* = 215 K and up to 300 K, the
$\rho$(T) dependence is linear with a slope $a$ = $d\rho$/$dT$ =
2.050 $\mu\Omega\cdot$ cm/K. The slope was calculated by fitting the
experimental curve using a computer and confirmed the linear
behavior of $\rho$(T) with a mean-root-square error of 0.009 $\pm$
0.002 in the specified temperature range. The pseudogap opening
temperature $T$* >> $T$$_{c}$ was defined as a temperature at which
the resistive curve deviates downward from the linearity (Fig. 1).
The more precise approach to determine $T$* with accuracy $\pm$ 0.5
K is to explore the criterion [$\rho$(T) - $\rho_{0}$]/$aT$ = 1
\cite{deMello} (inset in Fig. 1), where, as before, $a$ designates
the slope of the extrapolated normal-state resistivity,
$\rho_{N}$(T), and $\rho_{0}$ is its intercept with the Y axis. Both
methods give the same $T$* = 215 K, which is typical for
well-structured YBCO films with $T$$_{c}$ $\approx$ 88 K and is in
good agreement with literature data \cite{SolDmReview,SolHab}.

\begin{figure}[h]
\begin{center}
\includegraphics[width=0.75\textwidth]{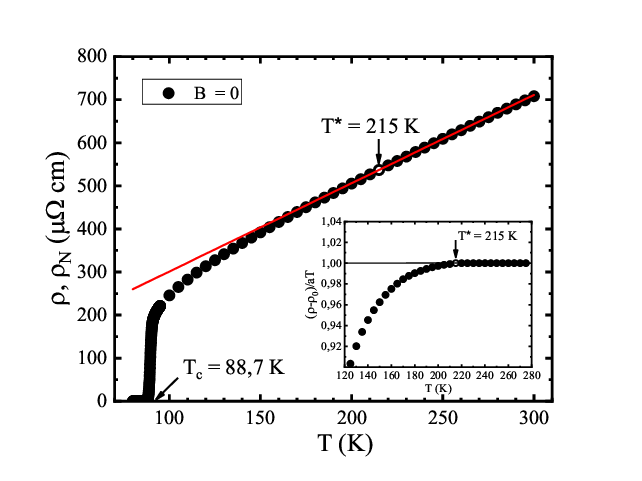}
\caption{$\rho$(T) dependence for YBa$_2$Cu$_3$O$_{7-\delta}$ film
in the absence of external magnetic field ($B$ = 0, dots). The solid
red line defines $\rho_{N}$(T), extrapolated to the low-temperature
region. The open circle corresponds to temperature $T$*. Inset:
method for determining $T$* using criterion [$\rho$(T) -
$\rho_{0}$]/$aT$ = 1 \cite{deMello}.}
\end{center}
\end{figure}

The influence of a magnetic field from $B$ = 0 to 8 T on the
temperature dependence $\rho$(T) is shown in Fig. 2. As can be seen
in the figure, the magnetic field noticeably broadens the resistive
transition, creating magnetoresistance, and reduces $T$$_{c}$ but,
as usual, does not affect the resistivity in the normal state
\cite{Oh,Nazarova}. It is worth noting the absence of any steps in
SC transitions in any magnetic field, which indicates the good
quality of the sample, its homogeneity, and the absence of
additional phases and inclusions. Unfortunately, for the 9 T case,
during the experiment, the contacts of the sample were damaged and
did not allow recording the temperature dependence in this field.

$\textbf{3.2. Excess conductivity}$\\

\begin{figure}[h]
\begin{center}
\includegraphics[width=0.75\textwidth]{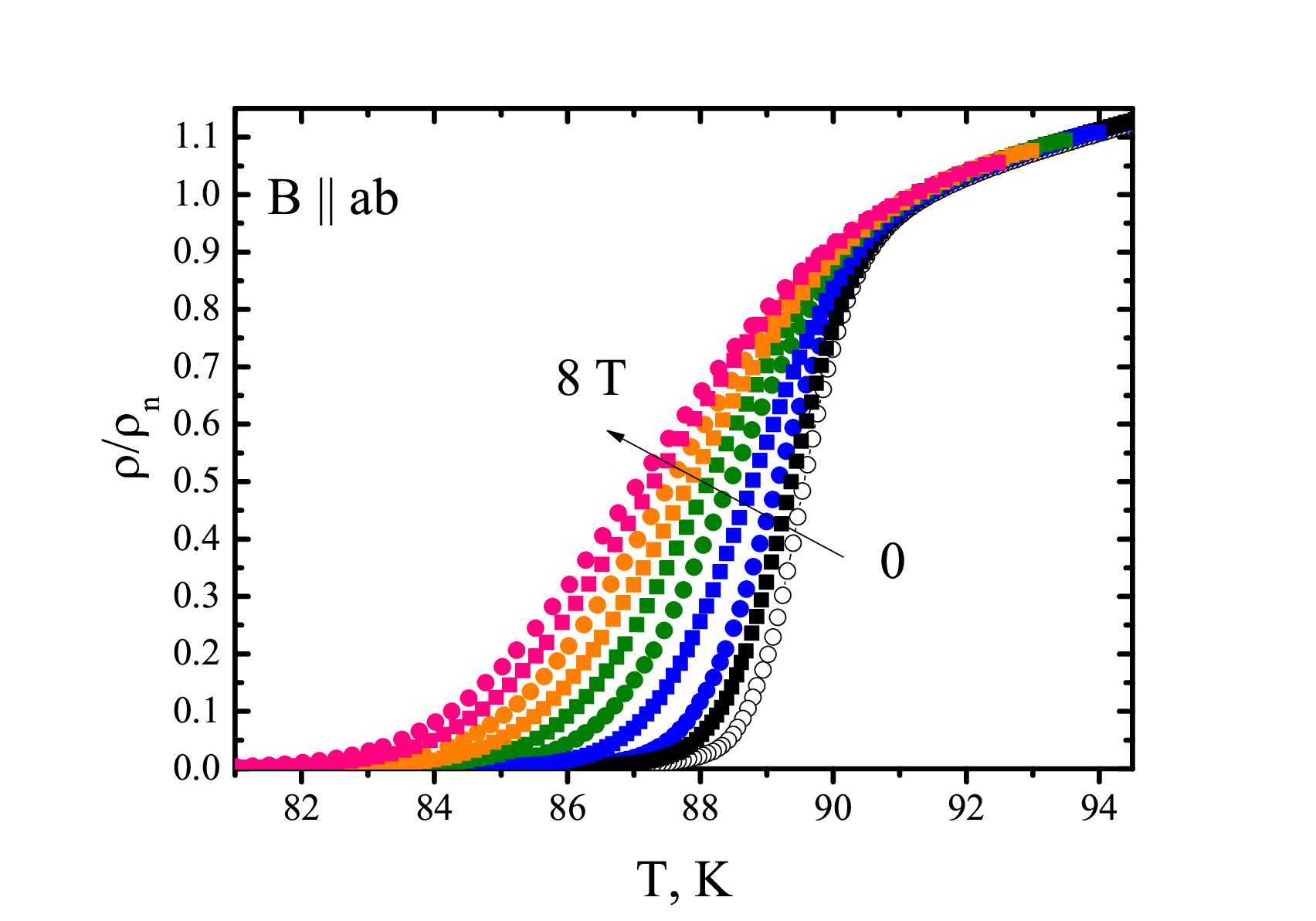}
\caption{Temperature dependences of $\rho$(T) in units of
($\rho$/$\rho_{n}$) of the YBa$_2$Cu$_3$O$_{7-\delta}$ film,
obtained for the field orientation parallel to the $ab$-plane ($B$
|| $ab$, $B$ = 0, 0.5, 1, 2, 3, 4, 5, 6, 7 and 8 T). $\rho_{n}$ =
203 $\mu\Omega\cdot$cm at $T$ = 92.3 K is a normal state resistivity
in the vicinity of the SC transition.}
\end{center}
\end{figure}

Below $T$* (Fig. 1), $\rho$(T) deviates from the linear dependence
toward smaller values. This leads to the appearance of excess
conductivity $\sigma'$(T), and the transition of HTSC to the PG
state (see Refs.\cite{Sol_rij,Pustovit,Lang} and references
therein). It is well known, that $\sigma'$(T) can be determined by
the simple formula \cite{SolDmReview,PetrExConMagF}:
\begin{equation}
\sigma '(T) = \sigma(T) -\sigma_N(T)= \frac{1}{\rho(T)}-
\frac{1}{\rho_N(T)}= \frac{\rho_N(T)-\rho(T)}{\rho(T)\cdot\rho_N(T)}
\label{sigma-t}
\end{equation}
therefore, it is actually determined by the measured resistivity
$\rho$(T) and the normal-state resistivity $\rho_{N}$(T)
extrapolated to the low $T$ region. This emphasizes that correctly
finding $\rho_{N}$(T) is of paramount importance for determining
$\sigma'$(T) and therefore $\Delta$*(T) \cite{Sol_rij}. In cuprates,
$\rho_{N}$(T) is a linear function of $T$ over a wide temperature
range above $T$* \cite{SolPhysB,Ando}. According to the NAFL (Nearly
Antiferromagnetic Fermi-liquid) model \cite{Stojkovic}, this linear
dependence corresponds to the normal state of HTSCs, characterized
by the stability of the Fermi surface. Below $T$*, numerous
anomalies of electronic properties are observed, associated with a
decrease in the density of single-particle excitations and
anisotropic rearrangement of the spectral density of charge
carriers, most likely due to the rearrangement of the Fermi surface
\cite{Badoux, Peng, Taillefer, Sol-Rog2023}. Finally, within the
framework of the local pair (LP) model \cite{Sol_rij}, based on
previously measured values of excess conductivity $\sigma'$(T), the
value and temperature dependence of the pseudogap parameter
$\Delta$*(T) was calculated for all applied magnetic fields \cite{PetrExConMagF}.\\

$\textbf{3.3. Field analysis of the pseudogap}$\\

Cuprates are known to have a pseudogap (PG) that opens below $T$*
and leads to the appearance of excess conductivity $\sigma'$(T) [Eq.
(1)], as mentioned above. Thus, it is believed that $\sigma'$(T)
should contain information about the magnitude and temperature
dependence of the pseudogap \cite{SolDmLP}. However, the question
remains open: how to extract this information, since there is still
no rigorous theory of HTSCs. As noted above, we share the idea that
PG in cuprates can be associated with the formation of local pairs
(LPs) at $T$ < $T$*, which at $T$ $\leq$ $T$* appear in the form of
SBBs subject to BEC, but with decreasing $T$ they gradually change
their properties to large fluctuating Cooper pairs (FCPs) near
$T_{c}$, obeying the BCS theory
\cite{Tallon,Esterlis,Yu,Sol_rij,Randeria,Emery,
Sol-Rog2023,SolDmLP}. The classical Aslamasov-Larkin (3D-AL)
\cite{AL} and Maki-Thompton (2D-MT) fluctuation theories, which were
modified by Hikami and Larkin (HL) \cite{HL} for HTSCs, provide a
good description of the experimental $\sigma'$(T) in cuprates, but
only in the range of SC fluctuations, that is, usually no more than
20 kelvins above $T_{c}$ \cite{SolPhysRevB}. Obviously, to obtain
information about the PG in the entire temperature range, from $T$*
to $T_{G}$, an equation is needed that describes the entire
experimental curve $\sigma'$(T) and contains the PG parameter
$\Delta$*(T) in explicit form. Such an equation was proposed in Ref.
\cite{SolDmLP} taking into account the LP model:
\begin{equation}
\sigma'(T)
=\frac{e^2A_4\left(1-\frac{T}{T^*}\right)\exp\left(-\frac{\Delta^*}{T}\right)}{16\hbar
\xi_{c}(0) \sqrt{2\varepsilon_{c0}^*
\sinh\left(\frac{2\varepsilon}{\varepsilon_{c0}^*}\right)}}
\label{sigma-t}.
\end{equation}
Here (1-$T$/$T$*) and exp(-$\Delta$*/$T$) take into account the
dynamics of LPs formation at $T$ $\leq$ $T$* and their destruction
by $kT$ near $T_{c}$, respectively.

Solving Eq. (2) with respect to $\Delta$*(T), we obtain
\begin{equation}
\Delta^*(T)
=T\text{ln}\frac{e^2A_4\left(1-\frac{T}{T^*}\right)}{\sigma'(T)16\hbar
\xi_{c}(0) \sqrt{2\varepsilon_{c0}^*
\sinh\left(\frac{2\varepsilon}{\varepsilon_{c0}^*}\right)}}
\label{sigma-t}
\end{equation}
where $\sigma'$(T) is the experimentally measured excess
conductivity over the entire temperature range from $T$* to $T_{G}$.
Accordingly, $A_{4}$ is a numerical coefficient that has the meaning
of the C-factor in the FLC theory, $\Delta$*($T$$_{G}$) is the value
of the PG parameter near $T_{c}$ and $\epsilon_{c0}^*$ is some
specific theoretical parameter that is discussed in detail below
\cite{Sol_rij,SolHab,SolDmLP,Oh}. As can be seen, equations (2) and
(3) contain a number of parameters that, importantly, can be
determined experimentally \cite{Sol_rij, SolPhysRevB, SolDmLP}. Such
parameters as $T$*, $T_{c}^{mf}$, reduced temperature $\epsilon$ =
($T$ - $T_{c}^{mf}$)/$T_{c}^{mf}$ and $\xi$$_{c}$(0) at various
values of $B$ have already been determined in our previous paper
devoted to the analysis of FLC in the same YBCO film
\cite{PetrExConMagF} (see Table I). $T_{c}^{mf}$ determines reduced
temperature and is therefore of primary importance for FLC and PG
analysis \cite{Oh, Sol_rij}. In fact, $T_{c}^{mf}$ is the mean-field
critical temperature, separating the range of SC fluctuations from
the critical fluctuations in the region of $T_{c}$, where the SC
order parameter $\Delta$ < $kT$ and Bogolyubov's mean-field theory
does not work \cite{GL,Lifshitz}. To find $T_{c}^{mf}$, we use the
fact \cite{Sol_rij,SolPhysB,SolPhysRevB,Sci.Rep.9-ybco_an} that in
all HTSCs, near $T_{c}$, $\sigma'$(T) is always described by the
standard equation of the 3D-AL theory \cite{AL}, in which
$\sigma'_{AL3D}$ $\sim$ $\epsilon^{-1/2}$ $\sim$ ($T$ -
$T_{c}^{mf}$)$^{-1/2}$. Accordingly, $\sigma'^{-2}$ $\sim$ ($T$ -
$T_{c}^{mf}$) and the intersection of its linear extrapolation with
the temperature axis just determines $T_{c}^{mf}$, since
$\sigma'^{-2}$ = 0 at $T$ = $T_{c}^{mf}$ \cite{Oh}. Note that always
$T_{c}^{mf}$ > $T_{c}$. The Ginzburg temperature $T_{G}$ >
$T_{c}^{mf}$ is another characteristic, down to which the mean-field
theory works \cite{Kapitulnik,Schneider}. The remaining parameters
shown in the table are discussed in detail in our previous article
\cite{PetrExConMagF}.
\begin{table}[tbp]
\caption [] \centering Parameters of FLC analysis of
YBa$_2$Cu$_3$O$_{7-\delta}$ film depending on the applied magnetic
field.
\begin{tabular}{|c|c|c|c|c|c|c|c|}
\hline $B$ & T$_\text{c}$ & T$_\text{c}^\text{mf}$ & T$_\text{G}$ &
T$_\text{0}$ & T$_\text{01}$ &
$\Delta$T$_\text{fl}$ & $\xi_\text{c}$(0)\\
(T) & (K) & (K) & (K) & (K) & (K) & (K) & (\AA) \\
[0.5ex] \hline 0 & 88.7 & 89.65 & 90.06 & 90.89 & 100.0 & 9.9 & 1.38 \\
[0.5ex] \hline 0.5 & 88.3 & 89.50 & 89.89 & 91.14 & 96.8 & 6.9 & 1.58\\
[0.5ex] \hline 1 & 87.9 & 89.40 & 89.89 & 91.12 & 94.5 & 4.6 & 1.62\\
[0.5ex] \hline 2 & 87.2 & 89.06 & 89.60 & 92.69 & 94.0 & 4.4 & 2.36\\
[0.5ex] \hline 3 & 86.6 & 88.88 & 89.51 & 100.0 & 120.0 & 30.5 & 4.13\\
[0.5ex] \hline 4 & 86.0 & 88.67 & 89.44 & 100.0 & 120.0 & 30.6 & 4.18\\
[0.5ex] \hline 5 & 85.6 & 88.46 & 89.24 & 100.0 & 115.0 & 25.8 & 4.22\\
[0.5ex] \hline 6 & 84.9 & 88.30 & 89.26 & 100.0 & 115.0 & 25.8 & 4.26\\
[0.5ex] \hline 7 & 84.5 & 88.09 & 89.10 & 100.0 & 115.0 & 25.9 & 3.99\\
[0.5ex] \hline 8 & 84.0 & 87.90 & 89.02 & 98.1 & 115.0 & 26.0 & 3.99\\
\hline
\end{tabular}
\label{tab:sample-values1}
\end{table}

As can be seen from the Table I, the magnetic field affects all
characteristic temperatures, except $T$* = 215 K, which actually
remains unchanged and is not shown. The same conclusions about the
independence of $T$* from the magnetic field were obtained both for
the YBCO \cite{Badoux} compounds and for the Bi$_2$
Sr$_2$CaCu$_2$O$_{8+y}$ (Bi-2212) \cite{TstarB, TstarBref}
compounds, despite the very strong magnetic fields used in the
experiment. These results once again emphasize the well-known fact
that even a strong magnetic field ($\sim$ 80 T, \cite{Badoux}) does
not have a noticeable effect on the resistivity of cuprates in the
normal state.

\begin{table}[tbp]
\caption [] \centering Parameters of the pseudogap analysis for
YBa$_2$Cu$_3$O$_{7-\delta}$ film depending on the applied magnetic
field.
\begin{tabular}{|c|c|c|c|c|c|c|c|}
\hline $B$ & $\varepsilon_{c0}^*$ & A$_\text{4}$ &
2$\Delta$*($T_{G}$)/$k$$_B$$T_{c}$ &
$T_{pair}$&$\Delta^*$($T_{pair}$)/$k_{B}$
&$\Delta^*$($T_{G}$)/$k_{B}$ & \textless
n$_{\uparrow}$n$_{\downarrow}$\textgreater\\

(T) & & & & (K) & (K) & (K) & \\
[0.5ex] \hline 0 & 0.26 & 7.7 & 5.0 & 140 & 254 & 219.5 & 0.307 \\
[0.5ex] \hline 0.5 & 0.28 & 9.3 & 5.0 & 140 & 254 & 218.3 & 0.305\\
[0.5ex] \hline 1 & 0.25 & 9.65 & 5.0 & 140 & 254 & 217.8 & 0.305\\
[0.5ex] \hline 2 & 0.25 & 15.3 & 5.0 & 135 & 254 & 215.3 & 0.301\\
[0.5ex] \hline 3 & 0.25 & 27.1 & 5.0 & 135 & 254 & 215.4 & 0.301\\
[0.5ex] \hline 4 & 0.25 & 27.8 & 5.0 & 135 & 254 & 213.4 & 0.299\\
[0.5ex] \hline 5 & 0.25 & 28.5 & 5.0 & 130 & 254 & 211.4 & 0.296\\
[0.5ex] \hline 6 & 0.25 & 29.1 & 5.0 & 130 & 254 & 209.8 & 0.294\\
[0.5ex] \hline 7 & 0.25 & 27.7 & 4.9 & 130 & 254 & 207.2 & 0.290\\
[0.5ex] \hline 8 & 0.25 & 28.0 & 4.9 & 130 & 254 & 203.8 & 0.285\\

\hline
\end{tabular}
\label{tab:sample-values1}
\end{table}

\begin{figure}[h]
\begin{center}
\includegraphics[width=1.00\textwidth]{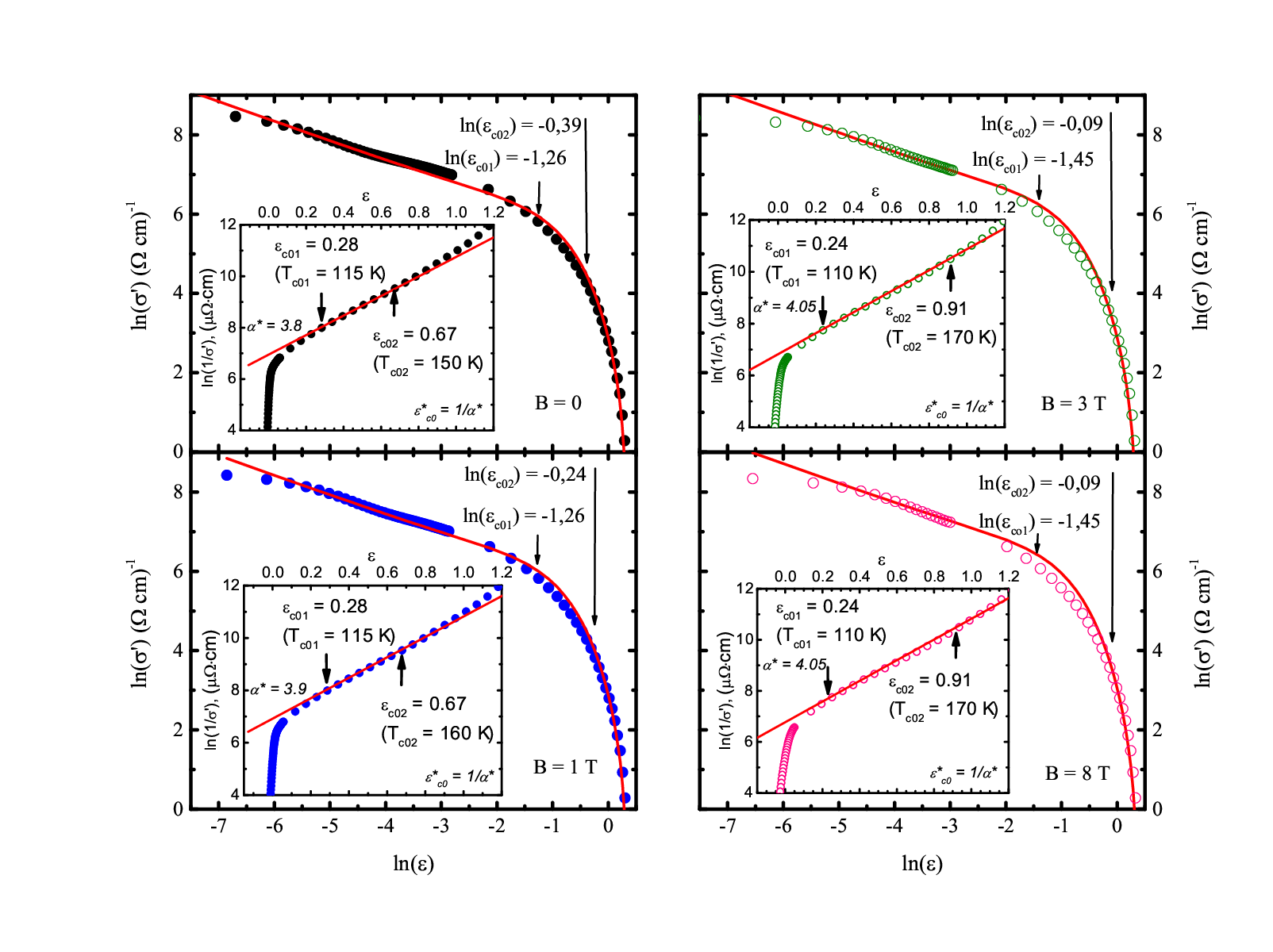}
\caption{ln$\sigma'$ vs ln$\epsilon$ (symbols) plotted in the entire
temperature range from $T$* down to $T_{c}^{mf}$ for $B$ = 0 (left
upper panel), 1 T (left lower panel), 3 T (right upper panel) and 8
T (right lower panel). The red solid curves at each panel are fits
to the data with Eq.(2). Insets: ln$\sigma'^{-1}$ as a function of
$\epsilon$. The straight red lines denote the linear parts of the
curves between $\epsilon_{c01}$ to $\epsilon_{c02}$. The slope
$\alpha^*$ determines the parameter $\epsilon_{c0}^*$ = 1/$\alpha^*$
(see text).}
\end{center}
\end{figure}

All missing parameters, required for Eq. 2 and 3, such as the
theoretical parameter $\epsilon_{c0}^*$, $\Delta$*($T_{G}$) and
coefficient $A_{4}$ can also be determined from experiment using the
approach developed within the LP model \cite{SolDmLP,SolDmReview}.
Fig. 3 shows the dependences of ln$\sigma'$ on ln$\epsilon$ for $B$
= 0, 1 T, 3 T and 8 T over the entire temperature range from $T$* to
$T_{G}$. It has been shown theoretically that $\sigma'^{-1}$ $\sim$
exp$\epsilon$ in a certain temperature range, indicated by the
arrows at ln$\epsilon_{c01}$ and ln$\epsilon_{c02}$ on the main
panels \cite{Leridon}. This feature turns out to be one of the main
properties of most HTSCs \cite{Sol_rij,SolPhysRevB,
SolDmLP,SolDmReview}. As a result, in the interval $\epsilon_{c01}$
< $\epsilon$ < $\epsilon_{c02}$ (see insets in panels),
ln$\sigma'^{-1}$ is a linear function of $\epsilon$ with a slope
$\alpha^*$, which determines the parameter $\epsilon_{c0}^*$ =
1/$\alpha^*$ \cite{Leridon}. This approach makes it possible to
obtain reliable values of $\epsilon_{c0}^*$ for all values of
applied magnetic field, which are given in Table II. Fig. 4
represents the same dependences for $B$ = 0 T (upper panel) and $B$
= 8 T (lower panel), but in coordinates ln$\sigma'$ vs 1/$T$. As
established in Refs. \cite{Sol_rij, SolPhysRevB,SolDmReview}, in
this case, the PG parameter $\Delta$*($T_{G}$) significantly affects
the shape of the theoretical curves presented in Fig. 4, at $T$ >
$T_{01}$, i.e., noticeably higher than the region of SC
fluctuations. Therefore, by selecting the best fit, the correct
value of $\Delta$*($T_{G}$) can be determined (Table II). In
addition, it was convincingly established that $\Delta$*($T_{G}$) =
$\Delta$(0), where $\Delta$ is the SC gap \cite{Yamada,Stajic}. We
emphasize that it is the value $\Delta$*($T_{G}$) that determines
the true value of PG and is used to estimate the value of the BCS
ratio 2$\Delta$(0)/$k$$_B$$T_{c}$ =
2$\Delta$*($T_{G}$)/$k$$_B$$T_{c}$ in a specific HTSC sample
\cite{SolPhysB, SolPhysRevB,SolDmReview}. The best approximation of
ln$\sigma'$ as a function of 1/$T$ from Eq. (2) for the case $B$ = 0
is achieved at 2$\Delta$*($T_{G}$)/$k$$_B$$T_{c}$ = 5.0 $\pm$ 0.05
(Table II), which corresponds to the strong coupling limit
characteristic of YBCO \cite{Inosov,Fischer, Dyachenko}.

\begin{figure}[h]
\begin{center}
\includegraphics[width=0.5\textwidth]{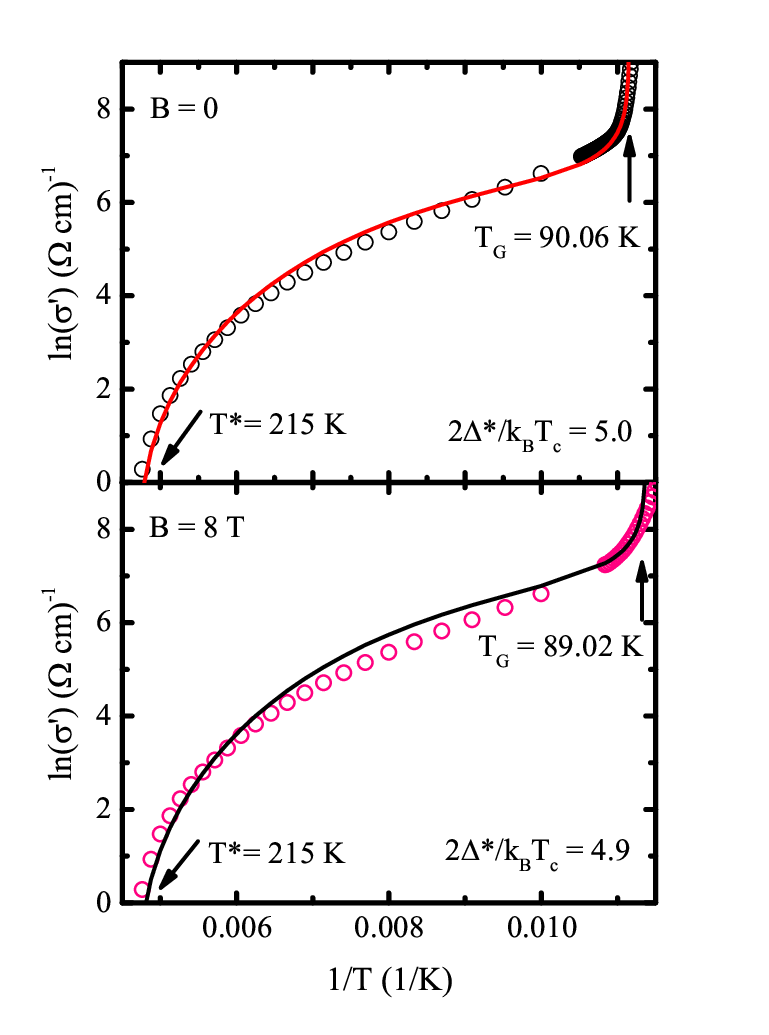}
\caption{ln$\sigma'$ vs 1/$T$ for $B$ = 0 (upper panel, black
circles) and 8 T (lower panel, pink circles) over the entire
temperature range from $T$* to $T_{G}$. The solid red and black
curves on each panel are fits to the data with Eq. (2). The best fit
is obtained when 2$\Delta$*($T_{G}$)/$k$$_B$$T_{c}$ = 5.0 $\pm$ 0.05
($B$ = 0) and 2$\Delta$*($T_{G}$)/$k$$_B$$T_{c}$ = 4.9 $\pm$ 0.05
($B$ = 8 T).}
\end{center}
\end{figure}

\begin{figure}[h]
\begin{center}
\includegraphics[width=1.00\textwidth]{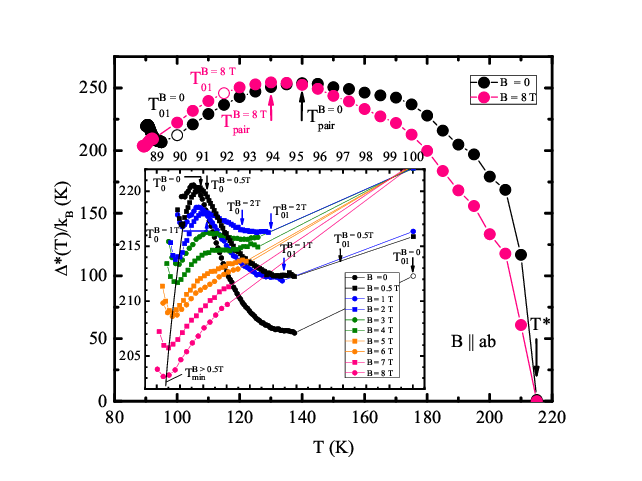}
\caption{$\Delta$*(T) as a function of $T$ of the studied
YBa$_2$Cu$_3$O$_{7-\delta}$ film, calculated by Eq. (3) for $B$ = 0
(black dots) and 8 T (pink dots). Empty circles indicate the
characteristic temperature $T_{01}$, which limits the range of SC
fluctuations from above (also in the inset for $B$ = 0). Inset: The
same dependences for the temperature interval $T_{G}$ < $T$ <
$T_{01}$ for $B$ = 0, 0.5, 1, 2, 3, 4, 5, 6, 7 and 8 T. The scales
in the inset have the same dimensions as the main figure. All
characteristic temperatures, both on the main panel and in the
insert, are indicated by arrows. The leftmost symbol of each curve
is $T_{G}$ that limits SC fluctuations from below. The auxiliary
black curve in the inset helps to trace the field evolution of the
low temperature minimum $T_{min}$ that appears at $B$ > 0.5 T. The
other solid curves are drawn to guide the eyes.}
\end{center}
\end{figure}

When $\epsilon_{c0}^*$ and $\Delta$*($T_{G}$) are known, the
coefficient $A_{4}$ can be determined by approximating the data
$\sigma'$($\epsilon$) by Eq. (2) and using the found parameters
(Tables I, II). With the correct choice of $A_{4}$ (Table II), the
theory is combined with experiment in the region of 3D AL
fluctuation near $T_{c}$, where
ln$\sigma'$($\epsilon$)(ln$\epsilon$) is a linear function of the
reduced temperature $\epsilon$ with slope $\lambda$ = -1/2
\cite{Sol_rij, SolPhysRevB,SolDmReview} (Fig. 3). As can be seen in
the figure, Eq. (2) describes well the experiment at temperatures
between $T$* and $T_{G}$ at $B$ = 0 as well as for all applied
magnetic fields. However, it is worth noting that above $\sim$ 2.5
T, in the temperature range from $T_{c01}$ to $T_{c02}$
(ln$\epsilon_{c01}$ and ln$\epsilon_{c02}$ in Fig. 3), the
experimental data deviate slightly downward from the theoretical
curve. This is most likely due to the fact that at $B$ > 2.5 T the
temperature dependence of the FLC changes from 2D MT to 2D AL
\cite{PetrExConMagF}, which is not taken into account in the LP
model.  At $B$ = 0 (left upper panel), the parameters are $A_{4}$ =
7.7, $\epsilon_{c0}^*$ = 0.26, and $\Delta$*($T_{G}$)/$k_{B}$ =
219.5 K (Table II). The values of the corresponding parameters for
all magnetic fields are also given in the Tables. As can be seen,
all parameters, except 2$\Delta$*($T_{c}$)/$k$$_B$$T_{c}$, change
noticeably with a change in the field, which suggests a possible
influence of the field on the PG as well.

The fact that all $\sigma'$($T,B$) are well described by equation
(2) (Fig. 3) suggests that equation (3) with the corresponding set
of found parameters should give reliable values, as well as
temperature and magnetic dependences of the parameter
$\Delta$*($T,B$) at any applied field. Figure 5 (main panel) shows
the result of the $\Delta$*($T$) analysis using Eq. (3) for $B$ = 0
and 8 T with all necessary for the PG analysis parameters given in
Tables I and II. As expected, at $B$ = 0 (black dots in the figure)
the shape of $\Delta$*($T$) is typical for YBCO films
\cite{SolDmLP,SolDmReview} and untwinned single crystals
\cite{Sci.Rep.9-ybco_an}, with a maximum at $T$ = $T_{pair}$
$\approx$ 140 K and a minimum at $T$ $\approx$ $T_{01}$
\cite{SolPhysRevB,ODYBCO1GPa,Sci.Rep.9-ybco_an}. As mentioned above,
$T_{pair}$ is the BEC-BCS crossover temperature
\cite{Emery,Randeria,Sol_rij,SolDmLP,Loktev,Tchernyshyov,Engel}.
Accordingly, as $T$ approaches $T_{c}$, there is a maximum of
$\Delta$*(T) just below $T_{0}$ and a minimum at $T$ = $T_{G}$
\cite{SolDmLP,SolDmReview} (see inset in Fig. 5). Below $T_{G}$,
there is usually a sharp jump in $\Delta$*(T) at $T$ $\rightarrow$
$T_{c}^{mf}$, which is deliberately not shown due to its
non-physical nature, since it corresponds to the transition to the
region of critical fluctuations, where the LP model does not work.
Thus, the approach in the framework of the LP model makes it
possible to determine the exact values of $T_{G}$ and, as a
consequence, to obtain reliable values of $\Delta$*$(T_{G})$
$\approx$ $\Delta$(0)
\cite{SolDmLP,SolDmReview,Leridon,Yamada,Stajic} depending on the
applied magnetic field (Table II).

\begin{figure}[h]
\begin{center}
\includegraphics[width=0.5\textwidth]{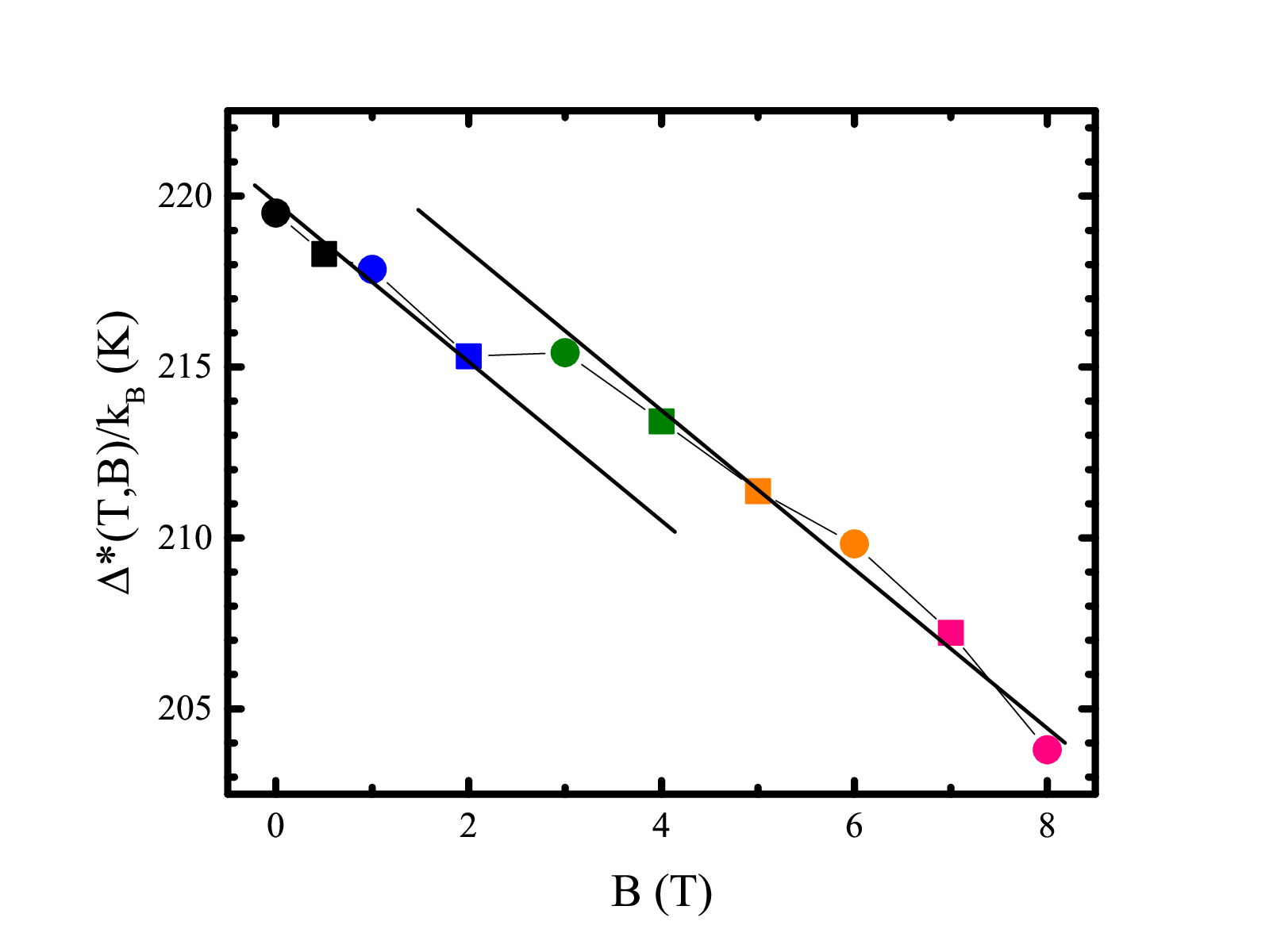}
\caption{$\Delta$*($T_{G}$) as a function of $B$, showing a fairly
good linear relationship, indicated by two dashed lines shifted by 1
T at $B$ > 2 T.  The thin solid line is drawn to guide the eyes.}
\end{center}
\end{figure}

Moreover, the $\Delta$*($T_{G}$) vs $B$ (Fig.6) can be approximated
by two linear sections with the same slope: the first one occurs
from 0 to 2 T and the second one, after a plateau between 2 and 3 T,
is clearly seen from 3 to 8 T. It should be noted, that the observed
behavior of $\Delta$*($T_{G}$) correlates with the FLC change from
2D MT to 2D AL \cite{PetrExConMagF}, as was already discussed above.

As can be seen from the Fig. 5 (main panel), a magnetic field $B$ =
8 T noticeably reduces the PG values at all temperatures above
$T_{pair}$, especially near $T$*. This seems quite surprising, since
the magnetic field has virtually no effect on the resistivity curve
at high temperatures (Fig. 1 and \cite{Matsuda,Oh}). Thus, our
result actually highlights the increased sensitivity of our approach
and leads us to the conclusion that the magnetic field has a much
stronger effect on local pairs than on normal electrons. However, to
explain this very important and somewhat unexpected result, it is
also necessary to keep in mind that, in accordance with the LP model
\cite{Sol-Rog2023}, the magnetic field, which is relatively weak in
our case, does not destroy the SSBs arising in HTSCs at $T$ < $T$*
\cite{Randeria}. On the other hand, we must also remember that PG
arises simultaneously with a decrease in the density of states (DOS)
at the Fermi level \cite{Alloul,Kondo}. This means that we can only
assume that the magnetic field somehow increases the DOS, thereby
leading to the observed decrease in the PG. It is possible that the
magnetic field somehow changes the distribution of normal electrons
and SBBs in the sample, thereby changing the DOS. However,
clarification of this issue requires special research.

From $T_{pair}$ down to $T_{01}$ the opposite dynamics is observed
(Fig. 5), expressed in a slight increase in $\Delta$*(T), which is
also surprising. Therefore, we can assume that the influence of the
magnetic field on the LPs and, possibly, on the DOS below $T_{pair}$
has also changed. Interestingly, the absolute value of
$\Delta$*($T_{pair}$) remains constant, while $T_{pair}$ shifts
towards low-temperature region from 140 K at $B$ = 0 to 130 K at 8
T. It seems that the entire $\Delta$*(T) curve also shifts towards
low temperatures along with the field, which can be considered as
another reason for the observed decrease in $\Delta$*(T). However,
measured decrease in $T_{c}^{mf}$  is $\Delta$$T_{c}^{mf}$ = 1.75K
and $\Delta$$T_{G}$ is only 1.04 K and, strictly speaking, is
clearly insufficient to account for the final shape of $\Delta$*(T).

At the same time, the BCS ratio 2$\Delta$*($T_{G}$)/$k$$_B$$T_{c}$
at $B$ > 7 T tends to decrease from 5.0 to 4.9 (Fig. 4 and Table
II). Note that all other film parameters, except
$\Delta$*($T_{pair}$) = 254 K, also change noticeably (Table II).
Our analysis of Eq. (3) showed that it is the change in the film
parameters that leads to the observed unexpected $\Delta$*(T) at $B$
= 8 T, since the value of the excess conductivity $\sigma'$(T) is
considered to be the same as at $B$ = 0. But, strictly speaking, the
physics of the change in the shape of PG under the influence of a
magnetic field remains unclear.

As $T$ approaches $T_{c}$, the magnetic field begins to intensively
reduce $\Delta$*(T) in the region of SC fluctuations below
$T_{01}$($B$ = 0) = 100 K (see inset in Fig. 5), while in a certain
temperature range up to $T_{pair}$($B$ = 8 T) = 103 K the
$\Delta$*(T) remains increased.  The evolution of $\Delta$*(T) below
$T_{01}$ with increasing magnetic field is as follows (see insert in
Fig. 5). The sharp low-temperature maximum near $T_{0}$ is
suppressed in magnitude and shifts toward higher temperatures and
completely disappears at $B$ > 5 T. At the same time, starting from
$\sim$ 0.5 T, above $T_{G}$ a pronounced minimum appears at
$T_{min}$. As can be seen in the insert, both $\Delta$*($T_{min}$)
and, more importantly, $\Delta$*($T_{G}$), which is the leftmost
point of each curve, decrease noticeably with increasing field,
simultaneously shifting towards lower temperatures. It is very
likely that this is due to the pair-breaking effect of the magnetic
field on the FCPs, leading to the appearance of noticeable
magnetoresistance, as well as the possible influence of the
two-dimensional vortex lattice created by the magnetic field, which
is discussed in \cite{PetrExConMagF}. However, from Fig. 6 it can be
seen that the decrease in $\Delta$*($T_{G}$) with increasing field
is also unusual. Indeed, the $\Delta$*($T_{G}$) dependence consists
of two linear sections with the same slope, but shifted by $\sim$ 1
T at $B$ > 2T. It should be noted, that the observed behavior of
$\Delta$*($T_{G}$) correlates with the change in FLC from the 3D-AL
to 2D-MT \cite{PetrExConMagF}, as discussed above. The revealed
behavior also correlates with the fact that all characteristic
temperatures detected in the analysis of FLC, as well as the range
of SC fluctuations $\Delta$$T_{fl}$ and the coherence length along
the $c$ axis $\xi_{c}$(0) increase sharply at $B$ $\ge$ 3 T (Table I
and \cite{PetrExConMagF}).  All these results suggest that pair
breaking is not the only factor in the influence of the magnetic
field.

The pair-breaking effect of the magnetic field should lead to a
decrease in the density of FCPs. To clarify this issue, we compare
the pseudogap parameters $\Delta$*(T)/$\Delta_{max}^*$, calculated
for all applied fields near $T_{c}$, with the Peters-Bauer (PB)
theory \cite{PB}. In the theory, within the framework of the
three-dimensional attractive Hubbard model, the temperature
dependences of the local density \textless
n$_{\uparrow}$n$_{\downarrow}$\textgreater\hspace{0pt} of pairs in
HTSCs are calculated for various temperatures $T/W$, interactions
$U/W$ and filling factors. In this case, $U$ is the activation
energy and $W$ is the band width. For all cuprates, the shape
$\Delta$*(T) near $T_{c}$, with a maximum near $T_{0}$ followed by a
minimum at $T_{G}$ \cite{SolPhysRevB} (see Fig. 12 in
\cite{SolPhysRevB} and Fig. 7 in \cite{ODYBCO1GPa}), resembles the
shape of theoretical \textless
n$_{\uparrow}$n$_{\downarrow}$\textgreater\hspace{0pt} curves at low
$T/W$ and $U/W$ \cite{PB}. This gives every reason to compare
experiment with the theory.

\begin{figure}[h]
\begin{center}
\includegraphics[width=1.00\textwidth]{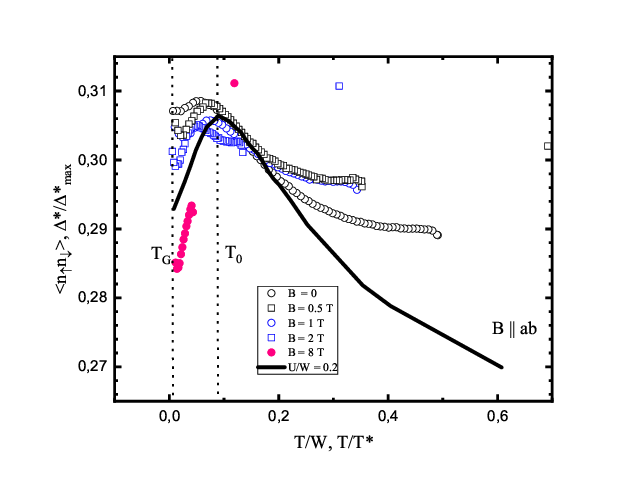}
\caption{$\Delta$*(T)/$\Delta_{max}^*$ as functions of $T/T$* near
$T_{c}$ for the magnetic fields $B$ = 0 (black circles), 0.5 (black
squares), 1 (blue circles), 2 (blue squares) and 8 T (pink dots)
compared with the theoretical dependence of \textless
n$_{\uparrow}$n$_{\downarrow}$\textgreater\hspace{0pt} on $T/W$ at
$U/W$ = 0.2 (black curve). The dashed lines indicate the
temperatures $T_{0}$ and $T_{G}$.}
\end{center}
\end{figure}

The comparison results are shown in Fig. 7. Symbols are the
experimental data near $T_{c}$ for $B$ = 0, 0.5, 1, 2 and 8 T. The
solid black curve corresponds to the theoretical dependence of
\textless n$_{\uparrow}$n$_{\downarrow}$\textgreater\hspace{0pt} on
$T/W$ at $U/W$ = 0.2. The dashed lines indicate the temperatures
$T_{0}$ and $T_{G}$. To carry out the analysis, the minimum
$\Delta$*(T)/$\Delta_{max}^*$ at $T_{G}$, measured at $B$ = 0, is
combined with the $T_{G}$ line, and the corresponding value
$\Delta$*(T)/$\Delta_{max}^*$ at $T_{0}$ is combined with the
$T_{0}$ line and then the result is compared with each theoretical
curve calculated at different $U/W$ values. The best fit is obtained
with a $U/W$ = 0.2 curve, indicating that in this case \textless
n$_{\uparrow}$n$_{\downarrow}$\textgreater\hspace{0pt} $\sim$ 0.3,
which is a typical value for various HTSCs at $B$ = 0
\cite{Sci.Rep.9-ybco_an, LTPfesePG, LTP49Ho}.  In this case, the
data fits the theoretical curve over the largest temperature range
above the maximum, as shown in the figure. This approach allows us
to find the approximation coefficients both along the X and Y axis,
which are then used at all values of the magnetic field.

As can be seen in Fig. 7, with an increase in the magnetic field,
the data at higher $T$ noticeably deviate from the theory upward,
and at $B$ = 8 T the data points (pink) do not fit the theoretical
curve at all. But more importantly, as the field increases, the
value of $\Delta$*(T)/$\Delta_{max}^*$ = \textless
n$_{\uparrow}$n$_{\downarrow}$\textgreater\hspace{0pt} $\approx$
0.307 at $B$ = 0 drops to \textless
n$_{\uparrow}$n$_{\downarrow}$\textgreater\hspace{0pt} $\approx$
0.285 at $B$ = 8 T. Simple algebra yields a reduction in local pair
density of about 7.0 \%.  However, if we now take the data shown in
the inset to Fig. 5, we will see that $\Delta$*(T) $\approx$ 219K at
$B$ = 0 and $\Delta$*(T) $\sim$ 204 K at $B$ = 8 T, that is, under
the influence of the field, $\Delta$*(T) decreases by approximately
the same 7.0 \%. Thus, we can conclude that near $T_{c}$, where the
LPs turn into FCPs, pair breaking by the magnetic field does play a
significant role in reducing PG. But the evolution of the shape of
$\Delta$*(T) near $T_{c}$ under the influence of a magnetic field
(Fig. 5) is most likely determined precisely by the dynamics of the
vortices created by the field in the range of SC fluctuations
and very likely through some other mechanisms \cite{PetrExConMagF}.\\

$\textbf{Conclusion}$\\

For the first time, within the framework of a local pair model, the
evolution of the temperature dependence of the pseudogap
$\Delta$*(T) in YBa$_2$Cu$_3$O$_{7-\delta}$ film with $T_{c}$ = 88.7
K under the influence of a magnetic field $B$ up to 8 T was studied
in detail. As expected, at $B$ = 0, $\Delta$*(T) in the entire
temperature range from the pseudogap opening temperature $T$* to
$T_{G}$, down to which the mean field theory works, has a wide
maximum at the BEC-BCS transition temperature $T_{pair}$  $\sim$ 140
K, which is typical for well-structured films and non-twinned single
crystals. Interestingly, despite the fact that the magnetic field
does not affect $\rho$(T) at high temperatures, all sample
parameters presented in the Tables change noticeably when the field
increases to 8 T. This leads to a noticeable change in the shape of
$\Delta$*(T), despite the fact that $T$* and excess conductivity are
considered unchanged. A shift in the temperature of the BEC-BCS
crossover $T_{pair}$ to the low temperature region from 140 K at $B$
= 0 to 130 K at 8 T was detected, while the maximum value
$\Delta$*($T_{pair}$) is maintained regardless of $B$. It looks like
the entire $\Delta$*(T) curve is somehow shifting towards lower
temperatures. The text discusses some attempts to describe the found
behavior of $\Delta$*(T), but, fairly speaking, the reason for the
unusual effect observed remains unclear.

In the region of SC fluctuations near $T_{c}$ the magnetic field
begins to intensively reduce $\Delta$*(T). The sharp low-temperature
maximum near $T_{0}$ is gradually suppressed, shifts toward higher
temperatures and completely disappears at $B$ > 5 T. At the same
time, starting from $\sim$ 0.5 T, above $T_{G}$, a pronounced
minimum appears at $T_{min}$. At the same time, both
$\Delta$*($T_{G}$), which is the leftmost point of each curve, and
$\Delta$*($T_{min}$) decrease noticeably with increasing field,
simultaneously shifting towards lower temperatures. The decrease in
$\Delta$*($T_{G}$) with increasing field turns out to be also
unusual. Indeed, the $\Delta$*($T_{G}$) dependence consists of two
linear sections with the same slope, but shifted by $\sim$ 1 T at
$B$ > 2 T. The revealed behavior also correlates with changes in
several other sample parameters at B $\ge$ 3 T, suggesting that pair
breaking is not the only factor in the influence of the magnetic
field.

The comparison of the data with the Peters-Bauer theory has shown
that near $T_{c}$, where the local pairs are transformed into
fluctuating Cooper pairs, pair breaking by the magnetic field does
play a significant role in reducing PG. But the specific evolution
of the shape $\Delta$*(T), discovered near $T_{c}$ under the
influence of a magnetic field, is most likely determined precisely
by the influence of the two-dimensional vortex lattice created by
the magnetic field, and also, probably, by some other mechanisms
that prevent the formation of superconducting fluctuations near $T_{c}$.\\

$\textbf{Acknowledgements}$\\

The work was supported by the National Academy of Sciences of
Ukraine within the F19-5 project.

We also acknowledge support from the European Federation of
Academies of Sciences and Humanities (ALLEA) though Grant
EFDS-FL1-32 (A.L.S.). A.L.S. also thanks the Division of Low
Temperatures and Superconductivity, INTiBS Wroclaw, Poland, for
their hospitality. Work was partly funded by Research Counsil of
Finland, project No.
343309.\\


\begin{thebibliography}{0}
\expandafter\ifx\csname natexlab\endcsname\relax\def\natexlab#1{#1}\fi
\expandafter\ifx\csname bibnamefont\endcsname\relax
  \def\bibnamefont#1{#1}\fi
\expandafter\ifx\csname bibfnamefont\endcsname\relax
  \def\bibfnamefont#1{#1}\fi
\expandafter\ifx\csname citenamefont\endcsname\relax
  \def\citenamefont#1{#1}\fi
\expandafter\ifx\csname url\endcsname\relax
  \def\url#1{\texttt{#1}}\fi
\expandafter\ifx\csname urlprefix\endcsname\relax\def\urlprefix{URL }\fi
\providecommand{\bibinfo}[2]{#2}
\providecommand{\eprint}[2][]{\url{#2}}

\end{thebibliography}


\begin{thebibliography}{54}

\bibitem{LK99} S. Lee, J.-H. Kim, Y.-W. Kwon, arXiv:2307.12008 (2023).

\bibitem{Garisto} D. Garisto, $Nature$ {\bf 620}, 705 (2023).

\bibitem{Puphal}P. Puphal, M. Y. P. Akbar, M. Hepting, E. Goering, M. Isobe, A. A. Nugroho, B. Keimer,  arXiv:2308.06256 (2023).

\bibitem{Kord} A. A. Kordyuk, \textit{Low Temp. Phys.} {\bf 41}, 319 (2015).

\bibitem{Gao} J. Gao, J. W. Park, K. Kim, S. K. Song, H. R. Park, J. Lee, J. Park, F. Chen, X. Luo, Y. Sun, and H. W. Yeom, \textit{Nano. Lett.} {\bf 20}, 6299-6305 (2020).

\bibitem{Tallon} J. L. Tallon, J. G. Storey, J. R. Cooper, and J. W. Loram, \textit{Phys. Rev. B} {\bf 101}, 174512 (2020).

\bibitem{Peng} Y. Y. Peng, R. Fumagalli, Y. Ding, M. Minola, S. Caprara, D. Betto, M. Bluschke, G. M. De Luca, K. Kummer, E. Lefran\c{c}is, M. Salluzzo, H. Suzuki, M. Le Tacon, X. J. Zhou, N. B. Brookes, B. Keimer, L. Braicovich, M. Grilli, and G. Ghiringhelli, \textit{Nat. Mater.} {\bf 17}, 697 (2018).

\bibitem{Chakraborty} D. Chakraborty, M. Grandadam, M. Y. Hamidian, J. C. S. Davis, Y. Sidis, and C.P\'epin, \textit{Phys. Rev. B} {\bf 100}, 224511 (2019).

\bibitem{Esterlis} I. Esterlis, S. A. Kivelson, and D. J. Scalapino, \textit{Phys. Rev. B} {\bf 99}, 174516 (2019).

\bibitem{Yu} G. Yu, D.-D. Xia, D. Pelc, R.-H. He, N.-H. Kaneko, T. Sasagawa, Y. Li, X. Zhao, N. Barisic, A. Shekhter, and M. Greven, \textit{Phys. Rev. B} {\bf 99}, 214502 (2019).

\bibitem{Loktev} V. M. Loktev, R. M. Quick, and S. G. Sharapov, \textit{Phys. Rep.} {\bf 349}, 1 (2001).

\bibitem{Sol_rij} A. L. Solovjov, \textit{Pseudogap and local pairs in high-T$_{c}$ superconductors, in: Superconductors-Materials, Properties and Applications}, A. Gabovich (ed.), InTech, Rijeka (2012), Chap. 7, p. 137.

\bibitem{Haussmann} R. Haussmann, \textit{Phys. Rev. B} {\bf 49}, 12975 (1994).

\bibitem{Tchernyshyov} O. Tchernyshyov, \textit{Phys. Rev. B} {\bf 56}, 3372 (1997).

\bibitem{Engel} J. R. Engelbrecht, A. Nazarenko, M. Randeria, and E. Dagotto, \textit{Phys. Rev. B} {\bf 57}, 13406 (1998).

\bibitem{Emery} V. J. Emery and S. A. Kivelson, \textit{Nature} {\bf 374}, 434 (1995).

\bibitem{Randeria} M. Randeria, \textit{Nat. Phys.} {\bf 6}, 561 (2010).

\bibitem{Sol-Rog2023} A. L. Solovjov and K. Rogacki, \textit{Low Temp. Phys} {\bf 49}, 375 (2023).

\bibitem{SolDmReview} A. L. Solovjov, V. M. Dmitriev, \textit{Low Temp. Phys.} {\bf 35}, 169 (2009).

\bibitem{Stojkovic} B. P. Stojkovi\'c, and D. Pines, \textit{Phys. Rev. B} {\bf 55}, 8576 (1997).

\bibitem{Badoux} S. Badoux et al., \textit{Nature (London)} {\bf 531}, 210 (2016).

\bibitem{Gabovich} A. M. Gabovich and A. I. Voitenko, \textit{Low Temp. Phys.} {\bf 42}, 1103 (2016).

\bibitem{Taillefer} L. Taillefer, \textit{Annu. Rev. Condens. Matter Phys.} {\bf 1}, 51 (2010).

\bibitem{Dzhumanov1} S. Dzhumanov, U.T. Kurbanov \textit{Modern Physics Letters B} {\bf 32}, 1850312 (2018).

\bibitem{Dzhumanov2} S. Dzhumanov, U.T. Kurbanov \textit{Superlattices and Microstructures} {\bf 84}, 66 (2015).

\bibitem{Wang} S. Wang, P. Choubey, Y. X. Chong, W. Chen, W. Ren, H. Eisaki, S. Uchida, P. J. Hirschfeld and J. C. S. Davis,  \textit{Nature Com.} {\bf12}, 6087 (2021).

\bibitem{Kivelson} S. F. Kivelson, and S. Lederer, \textit{PNAS} {\bf 116}, 14395 (2019).

\bibitem{Robinson} N. J. Robinson, P. D. Johnson, T. M. Rice, and A. M. Tsvelik, \textit{Rep. Prog. Phys.} {\bf 82}, 126501 (2019).

\bibitem{Mishra} V. Mishra, U. Chatterjee, J.C. Campuzano,  and M.R. Norman, \textit{Nat. Phys.} {\bf 10}, 357 (2014).

\bibitem{VovkExConMagF} R. V. Vovk, Z. F. Nazyrov, G. Ya. Khadzhai, V. M. Pinto Simoes, V. V. Kruglyak, \textit{Functional Materials} {\bf 20}, 208 (2013).

\bibitem{Malik} B. A. Malik, G. H. Rather, K. Asokan, M. A. Malik, \textit{Applied Physics A} {\bf 127} (2021).

\bibitem{Rey} R. I. Rey, C. Carballeira, J. M. Doval, J. Mosqueira, M. V. Ramallo, A. Ramos-\'Alvarez, D. S\'o\~nora1, J. A. Veira, J. C. Verde and F. Vidal, \textit{Supercond. Sci. Technol.} {\bf 32}, 045009 (2019).

\bibitem{PetrExConMagF} E. V. Petrenko, L. V. Omelchenko, Yu. A. Kolesnichenko, N. V. Shytov, K. Rogacki, D. M. Sergeyev, and A. L. Solovjov, \textit{Low Temp. Phys.} {\bf 47}, 1148 (2021).

\bibitem{DeGennes} P. G. De Gennes, \textit{Superconductivity of Metals and Alloys} W. A. Benjamin, Inc., New York, Amsterdam (1968) p. 280.

\bibitem{ODYBCO1GPa} A. L. Solovjov, L. V. Omelchenko, R. V. Vovk, O. V. Dobrovolskiy, S. N. Kamchatnaya, D. M. Sergeev, \textit{Curr. Appl. Phys.} {\bf 16}, 931 (2016).

\bibitem{AL} L. G. Aslamazov and A. L. Larkin, \textit{Phys. Lett. A} {\bf 26}, 238 (1968).

\bibitem{HL} S. Hikami and A. I. Larkin, \textit{Mod. Phys. Lett. B} {\bf 2}, 693 (1988).

\bibitem{Sci.Rep.9-ybco_an} A. L. Solovjov, E. V. Petrenko, L. V. Omelchenko, R. V. Vovk, I. L. Goulatis, and A. Chroneos, \textit{Sci. Rep.} {\bf 9}, 9274 (2019).

\bibitem{PB} R. Peters and J. Bauer, \textit{Phys. Rev. B} {\bf 92}, 014511 (2015).

\bibitem{Przyslupski} P. Przyslupski, I. Komissarov, W. Paszkowicz, P. Dluzewski, R. Minikayev, M. Sawicki, \textit{J. Appl. Phys.} {\bf 95}, 2906 (2004).

\bibitem{deMello} E. V. L. de Mello, M. T. D. Orlando, J. L. Gonzalez, E. S. Caixeiro, and E. Baggio-Saitovich, \textit{Phys. Rev. B} {\bf 66}, 092504 (2002).

\bibitem{SolHab} A. L. Solovjov, H. - U. Habermeier, T. Haage, \textit{Low Temp. Phys.} {\bf 28}, 144 (2002).

\bibitem{Oh} B. Oh, K. Char, A. D. Kent, M. Naito, M. R. Beasley, T. H. Geballe, R. H. Hammond, A. Kapitulnik, and J. M. Graybeal, \textit{Phys. Rev. B} {\bf 37}, 7861 (1988).

\bibitem{Nazarova} E. Nazarova, A. Zaleski, K. Buchkov, \textit{Physica C} {\bf 470}, 421 (2010).

\bibitem{Pustovit} Y. V. Pustovit and A. A. Kordyuk, \textit{Fiz. Nizk. Temp.} {\bf 42}, 1268 (2016) [\textit{Low Temp. Phys.} {\bf 42}, 995 (2016)].

\bibitem{Lang} W. Lang, G. Heine, P. Schwab, X. Z. Wang, and D. Bauerle, \textit{Phys. Rev. B} {\bf 49}, 4209 (1994).

\bibitem{SolPhysB} A. L. Solovjov, L. V. Omelchenko, R. V. Vovk, O. V. Dobrovolskiy, Z. F. Nazyrov, S. N. Kamchatnaya, and D. M. Sergeyev, \textit{Physica B} {\bf 493}, 58 (2016).

\bibitem{Ando} Y. Ando, S. Komiya, K. Segawa, S. Ono, and Y. Kurita, \textit{Phys. Rev. Lett.} {\bf 93}, 267001 (2004).

\bibitem{SolDmLP} A. L. Solovjov and V. M. Dmitriev, \textit{Fiz. Nizk. Temp.} {\bf 32}, 139 (2006) [\textit{Low Temp. Phys.} {\bf 32}, 99 (2006)].

\bibitem{SolPhysRevB} A. L. Solovjov, L. V. Omelchenko, V. B. Stepanov, R. V. Vovk, H.-U. Habermeier, H. Lochmajer, P. Przyslupski, and K. Rogacki, \textit{Phys. Rev. B} {\bf 94}, 224505 (2016).

\bibitem{GL} V. L. Ginzburg and L. D. Landau, "On the theory of superconductivity", in \textit{On Superconductivity and Superfluidity} (Springer, Berlin, Heidelberg, 2009).

\bibitem{Lifshitz} E. M. Lifshitz and L. P. Pitaevski, \textit{Statistical Physics} (Nauka, Moscow, 1978).

\bibitem{Kapitulnik} A. Kapitulnik, M. R. Beasley, C. Castellani, and C. Di Castro, \textit{Phys. Rev. B} {\bf 37}, 537 (1988).

\bibitem{Schneider} T. Schneider and J. M. Singer, \textit{Phase Transition Approach to High-Temperature Superconductivity: Universal Properties of Cuprate Superconductors} (Imperial College Press, London, 2000).

\bibitem{TstarB} P. Pieri, G. C. Strinati, and D. Moroni, \textit{Phys. Rev. Lett.} {\bf 89}, 127003 (2002).

\bibitem{TstarBref} T. Shibauchi, L. Krusin-Elbaum, Ming Li, M. P. Maley, and P. H. Kes, \textit{Phys. Rev. Lett.} {\bf 86}, 5763 (2001).

\bibitem{Leridon} B. Leridon, A. Defossez, J. Dumont, J. Lesueur, and J. P. Contour, \textit{Phys. Rev. Lett.} {\bf 87}, 197007 (2001).

\bibitem{Yamada} Y. Yamada, K. Anagawa, T. Shibauchi, T. Fujii, T. Watanabe, A. Matsuda, and M. Suzuki, \textit{Phys. Rev. B} {\bf 68}, 054533 (2003).

\bibitem{Stajic} J. Stajic, A. Iyengar, K. Levin, B. R. Boyce, and T. R. Lemberger, \textit{Phys. Rev. B} {\bf 68}, 024520 (2003).

\bibitem{Inosov} D. S. Inosov, J. T. Park, A. Charnukha, Y. Li, A. V. Boris, B. Keimer, and V. Hinkov, \textit{Phys. Rev. B} {\bf 83}, 214520 (2011).

\bibitem{Fischer} \O. Fischer, M. Kugler, I. Maggio-Aprile, and C. Berthod, \textit{Rev. Mod. Phys.} {\bf 79}, 353 (2007).

\bibitem{Dyachenko} A. I. D'yachenko, V. Y. Tarenkov, V. V. Kononenko, E. M. Rudenko, \textit{Metallofiz. Noveishie Tekhnol.} {\bf 38}, 565 (2016).

\bibitem{Matsuda} Y. Matsuda, T. Hirai, S. Komiyama, T. Terashima, Y. Bando, K. Iijima, K. Yamamoto, and K. Hirata, \textit{Phys. Rev. B} {\bf 40}, 5176 (1989).

\bibitem{Alloul} H. Alloul, F. Rullier-Albenque, B. Vignolle, D. Colson, and A. Forget, \textit{EPL} {\bf 91}, 37005 (2010).

\bibitem{Kondo} T. Kondo, A. D. Palczewski, Yoichiro Hamaya, T. Takeuchi, J. S. Wen, Z. J. Xu, G. Gu, and A. Kaminski, \textit{Phys. Rev. Lett.} {\bf 111}, 157003 (2013).

\bibitem{LTPfesePG} A. L. Solovjov, E. V. Petrenko, L. V. Omelchenko, E. Nazarova, K. Buchkov, and K. Rogacki, \textit{Fiz. Nizk. Temp.} {\bf 46}, 638 (2020) [\textit{Low. Temp. Phys.} {\bf 46}, 538 (2020)].

\bibitem{LTP49Ho} A. L. Solovjov, L. V. Omelchenko, E. V. Petrenko, Yu. A. Kolesnichenko, A. S. Kolesnik, S. Dzhumanov, and R. V. Vovk, \textit{Low. Temp. Phys.} {\bf 49}, 115 (2023).

\end{thebibliography}
\end{document}